\documentclass{ws-p8-50x6-00}

\begin{document}

\title{What Can We Learn from Charmless Rare~B~Decays 
	--- the Past/Next 3 Years}

\author{George W.S. Hou}

\address{Department of Physics, National Taiwan University,
Taipei, Taiwan, R.O.C. 
}

\maketitle

\abstracts{
A personal perspective is given on 
physics of charmless rare B decays: 1997 -- 2003.}

\def\ltap{\ \raisebox{-.5ex}{\rlap{$\sim$}} \raisebox{.4ex}{$<$}\ }
\def\gtap{\ \raisebox{-.5ex}{\rlap{$\sim$}} \raisebox{.4ex}{$>$}\ }

\section{Prelude: Berkeley to Osaka}

My first ICHEP was Berkeley 1986, where I heard about CLEO limits
on rare B decays. 
A host of strong (P, e.g. $B\to \phi K^+$), 
radiative or electromagnetic (EMP, e.g. $B\to K^*\gamma$) and
electroweak (EWP, e.g. $B\to K\ell^+\ell^-$) modes
were given, with limits typically a few $\times 10^{-4}$,
but EMP limits were only $\sim 10^{-3}$,
reflecting the time before the advent of CsI-based EM calorimetry.
\footnote{
The first penguin, $B\to K^*\gamma$, 
emerged $\sim 5 \times 10^{-5}$ soon
after CLEO-II (w/ CsI) started.}

Though still contending with limits,
I embarked on a study of these modes,
encountering (and missing) a few surprises.
I found that the $Z$ penguin dominated
the EWP's.\cite{HWS}
While obvious for $b\to s\nu\bar\nu$,
for $b\to s\ell^+\ell^-$ the EMP had naively
been assumed to be dominant.
Although I missed the ``large QCD corrections" to EMP 
that enhanced $b\to s\gamma$ to a few $\times 10^{-4}$,
I did uncover the sensitivity\cite{HW} to $H^+$ effect,
which later lead to the stringent bound on $m_{H^+}$
in SUSY type of two Higgs models.
I encountered the surprise of ``higher order dominance"
that\cite{HSS} $b\to sg^* \to s\bar qq \ > \ b\to sg$.
Much of the above has to do with GIM cancellation subtleties
or the non-decoupling of $m_t$ in SM.

Why did I bother only with inclusive processes at that time?
Well, these can be, by argument of duality,
viewed as quark level processes hence 
dominated by S.D. physics 
($\sim$ virtual collider).
But for {\it exclusive} processes,
hadronization brings in L.D. physics that
is likely to blur S.D. information.
Unless, \ldots ugh, \ldots one has {\it factorization!?}
Factorization means
separating the product of quark currents
and forming hadronic matrix elements of bilinears.
Naively, for $b\to d\bar uu$ to mediate $\bar B\to \pi^-\pi$,
the $\bar du$ current created by virtual $W$
projects directly into a pion (decay constant),
while the $b\to u$ current 
mediates $B\to \pi$ transition (form factor).
That is, one {\it assumes} ({\it wishes})
\[
\langle\pi\pi\vert(\bar du)(\bar ub)\vert B\rangle
\stackrel{\rm fac.}{\sim}
\langle\pi\vert\bar du\vert 0\rangle 
\langle\pi\vert\bar ub\vert B\rangle.
\]
But this was too big an ``if" for me.
Despite the ease for detection,
I could not bring myself to believe in factorization,
and hence never worked on exclusive modes
\ldots until I had a {\it data--driven} 
conversion experience at end of 1998!

Strong penguins have emerged {\it en masse} since 1997 at
CLEO, where $10^7$ $B\bar B$'s have been collected, 
and at Osaka 2000,
Belle and BaBar both reported first results
with comparable amounts of data,
{\it collected in first year of running!}
The four $K\pi$ modes 
and relatively small $\pi^+\pi^-$ mode observed by CLEO
are confirmed by Belle,
but BaBar reports a higher (lower) $\pi^+\pi^-$ ($K^+\pi^-$).
More data is needed, and certainly expected,
but Golutvin\cite{Gol} already states: 
``Present data favour large $\arg V_{ub}$ and FSI."
We shall address why this is so,
but let me step a few years back to 
the time of the emerging strong penguins.

\section{Experimental Surprise of $\eta^\prime X_s$
	 and $\eta^\prime K$}

With a few $\times 10^{6}$ $B\bar B$'s,
CLEO announced in 1997 that 
$B\to \eta^\prime X_s \sim 6\times 10^{-4}$
for $2 < p_{\eta^\prime} < 2.7$ GeV,
and $\eta^\prime K$ a tenth less.
{\it This was not predicted by any theorist}.
Many models and speculations ensued.
It is puzzling that,
with $\eta^\prime$ the heaviest and stickiest (gluey)
member of the $0^-$ nonet,
how can it come so ``fast"?
One interesting idea,\cite{AS,HT} the only one so far that can
explain the $m_{X_s}$ spectrum,
links $\eta^\prime$ production to the gluon anomaly:
The derivative coupling nature of
the $g^*$-$g$-$\eta^\prime$ anomaly vertex
spits out $\eta^\prime$ with high momentum
in $b\to sg^* \to sg\eta^\prime$.
A criticism is that
such coupling must be cut off by some form factor.
However, the $g^*g$ channel has $0^{-+}$ quantum numbers,
and the high glueball mass scale ($\sim 2.5$ GeV from lattice)
may well delay the form factor suppression.
One way\cite{HT}
to check this is to study $Z\to q\bar q g + \eta^\prime$.

The $\eta^\prime K$ analysis of CLEO is now very robust, 
and is starting to be checked by Belle/BaBar,
%
but inclusive study has not yet been updated 
by any group (even CLEO).
This is certainly a volatile area
where more insight, if not surprises,
can be attained.
We still lack a clear theory.

\section{The Path to $\gamma > 90^\circ$ and Factorization
	 (and FSI?)
}

\noindent\underline{\bf Factorization and \boldmath{$\gamma \gtap 90^\circ$}}
\vskip0.1cm

CLEO data has driven phenomenology in a fine way in 
the last 3 years.\cite{Osaka}
\begin{description}
\item[1997]: $\bar K^0\pi^- > K^-\pi^+$
----- This lead to the Fleischer--Mannel bound 
(though $\bar K^0\pi^-$ was just above 3$\sigma$),
and a boom in theory work,
eventually leading to model-independent methods 
for extracting $\gamma$.
\item[1998]: $K^-\pi^0 \gtap \bar K^0\pi^- \simeq K^-\pi^+
	 \simeq 1.4\times 10^{-5}$
----- $K^0\pi \simeq K\pi$ prompted 
the first suggestion\cite{DHHP} for large $\gamma$;
surprising strength of $K\pi^0$ indicated EWP.
\item[1999]: $\rho^0\pi^-$, $\rho^\pm\pi^\mp$, $\omega\pi$
	 ($\omega K$ disappear)
	  ----- Evidence for $b\to u$ tree (T).\\ 
\phantom{Xi}
$K\pi^0\simeq {2\over 3}\; (\bar K^0\pi\simeq K\pi)$
 ----- Fine, EWP at work.\cite{DHHP}\\
\phantom{Xi}
$\pi\pi \sim {1\over 4}\; K\pi$
 ----- Further indication for large $\gamma$.\cite{HHY}\\
\phantom{Xi}
$K^0\pi^0 \sim K\pi,\ K^0\pi \Longrightarrow$ \underline{Problem}.\cite{HY}
\end{description}

The host of emerging modes lead to the observation\cite{HHY} that,
\footnote{
Only the sign change in $\cos\gamma$ was 
conservatively advocated in original paper.
}
\[
\fbox{\it Factorization\ {\it works}\ in\ 
two\ body\ charmless\ rare\ B\ decays,\
{\bf if}\ $\cos\gamma \ltap 0$}
\]
It even lead to\cite{HSW} a 
``global (rare B) fit" of more than 10 modes
that gave $\gamma \simeq 105^\circ$,
seemingly in some conflict with
the well-known ``CKM Fit" value of $\gamma\simeq 60^\circ$.
Sufficie it to say that, by end of 1999,
all B practioners had switched to 
$\gamma \gtap 80^\circ$-$90^\circ$,
as reflected in the 5 rare B theory talks at Osaka.
This could have harbingered 
the lower central value of $\sin2\beta$ seen by BaBar/Belle
this summer, although dust is far from settled.

\vskip-0.5cm
\begin{figure}[htb]
\centerline{\hskip-1.6cm
            {\epsfxsize3.0 in \epsffile{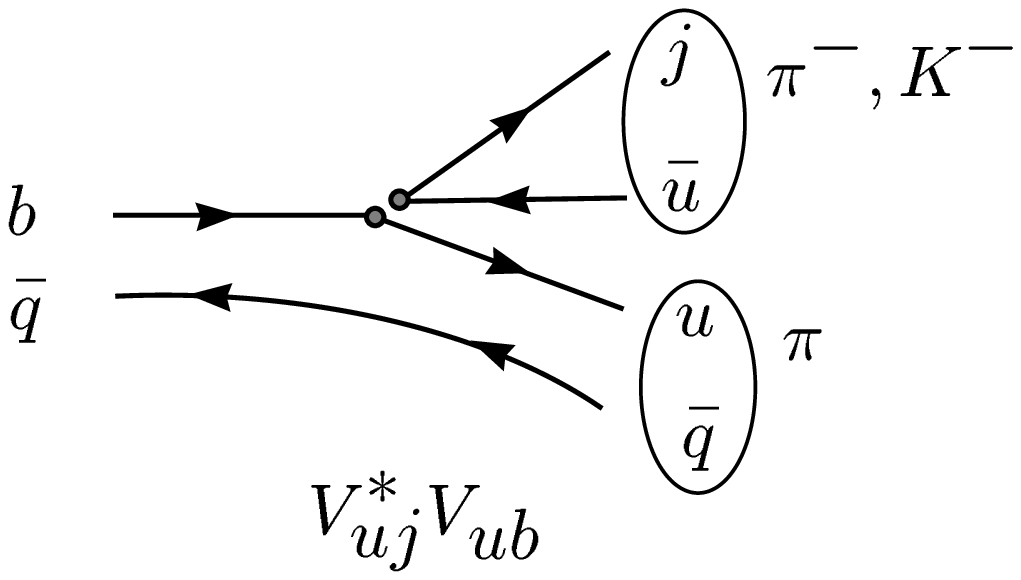}} \hskip-1.9cm
            {\epsfxsize3.0 in \epsffile{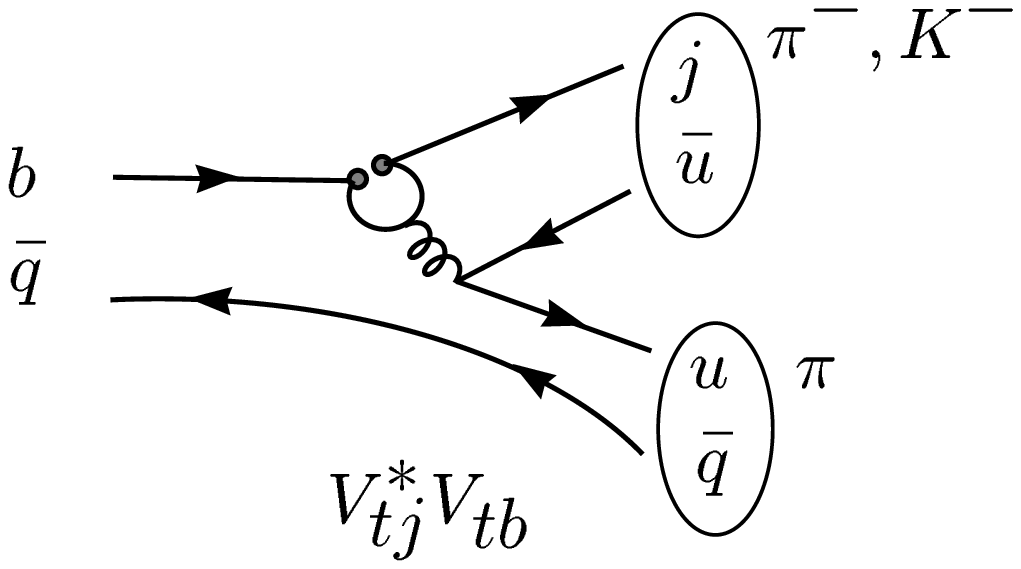}}
	     }
\vskip-2.0cm
\end{figure}

%
What is the physics effect?
Let us illustrate with $K\pi$ vs. $\pi\pi$. 
With only T contributions,
$\vert V_{us}\vert^2 \ll \vert V_{ud}\vert^2$ implies $K\pi \ll \pi\pi$.
Thus, the observed $K\pi \gtap \pi\pi$ implies P dominance in $K\pi$,
and substantial ``penguin pollution" to $\pi\pi$.
As data refined, it was realized\cite{HHY} that 
T-P interference contains more information.
As 
$V_{us}^*V_{ub} = \lambda\vert V_{ub}\vert e^{-i\gamma}
                = \lambda V_{ud}^*V_{ub}$
has the same phase,
while the real part of $V_{ts}^*V_{tb} \cong -\vert V_{cb}\vert$ 
and $V_{td}^*V_{tb} \cong \lambda \vert V_{cb}\vert
[1-\sqrt{\rho^2+\eta^2}e^{-i\gamma}]$
have opposite sign since $\sqrt{\rho^2+\eta^2} \equiv 
\vert V_{ub}\vert/\lambda \vert V_{cb}\vert \simeq 0.4$,
hence {\it T-P interference is anticorrelated in
$K\pi$ vs. $\pi\pi$}.
Thus, $K\pi$ $\Uparrow$ implies $\pi\pi$ $\Downarrow$,
and {\it vice versa}.\cite{HHY}

\vskip0.25cm
\noindent\underline{\bf FSI?}
\vskip0.1cm

A problem was already apparent by summer 1999:
$K^0\pi^0$ seems too large\cite{HY}
(Again a chorus line of theorists at Osaka)!
As mentioned, $K\pi^0/K\pi \simeq 0.65$
confirms {\it constructive}
EWP-P interference for $K\pi^0$ in SM.
From the operators and the $\pi^0$ w.f. 
(change from $u\bar u$ to $d\bar d$)
one expects {\it destructive} EWP-P interference
in $K^0\pi^0$, hence $K^0\pi^0 > K\pi^0$
is very hard to reconcile.

%
We proposed a half-way solution,
resorting to large final state interaction (FSI) phases.\cite{HY}
If we start with e.g. $\gamma = 110^\circ$,
then $K\pi : K^0\pi : K\pi^0 : K^0\pi^0 =
 1 : 0.94 : 0.65 : 0.35$, as compared to the experimental (CLEO only)
$1 : 1.06 : 0.67 : {\bf 0.85}$.
Allowing $\delta_{K\pi}$
(strong phase difference between $I = {3\over 2}$ and ${1\over 2}$ amplitudes)
to be $\sim 90^\circ$, we find the ratio becomes
$1 : 1.12 : 0.61 : 0.47$.
This is far from resolving the problem, 
but it is in the right direction.
What's more, we find that $\pi\pi < \pi\pi^0$ can be achieved
(taking $\delta_{\pi\pi} \sim \delta_{K\pi}$)
and central values for $a_{\rm CP}$ in
$K\pi$, $K^0\pi$, $K\pi^0$ modes become ``just right",
and there are further dramatic consequences:

\noindent \phantom{x} $\bullet$
 $\pi^0\pi^0 \sim \pi\pi \sim 3$--$5 \times 10^{-6}
	 \ltap \pi\pi^0$ (still satisfy CLEO bound).

\noindent \phantom{x} $\bullet$
 $a_{\rm CP}^{\bar K^0\pi^0} \sim - a_{\rm CP}^{K^-\pi^0}$ large; \ \
	$a_{\rm CP}^{\pi\pi},\ a_{\rm CP}^{\pi^0\pi^0}$ as large as
	 $\sim -60\%$, $-30\%$ possible.
	 
\noindent {\it These would be measurable in a couple of years.}

So now we have an oxymoron: 
Factorization works, but FSI is large.
Our view is, however, phenomenological:
Data indicates that factorization
works for the first 10-20 or so two body rare B modes.
The $\delta_{K\pi}$ and $\delta_{\pi\pi}$ phases are 
the minimal extension of parameters allowed
in the factorization framework.
We do not pretend to know their origin.
They could be effective parameters arising from 
e.g. annihilation diagrams.
But if they genuinely arise from L.D. physics,
they would then pose a real problem for PQCD.

\section{New Physics: Probing Flavor and/or CP Violation}

As a ``virtual collider", B decays and mixings provide
a natural hunting ground for New Physics, esp.
flavor and CP violation. Let us illustrate with SUSY.

It is known that
$\tilde g$-$\tilde q$ loops could easily generate
$F_2^{\rm NP}\, \bar s\, i\sigma_{\mu\nu} m_b R\, b\, G^{\mu\nu}$
type couplings,
and also $b_L \to s_R$ chirality flips that are absent in SM.
For example, 
$a_{\rm CP}$ in inclusive $B\to \eta^\prime + X_s$ 
at 10\% level.\cite{HT}
A late 1997 rumor that CLEO had $a_{\rm CP}(K\pi) \sim 100\%$
led us to put in a sizable $F_2^{\rm NP}$
and managed\cite{HHY-NP} to yank $a_{\rm CP}$'s up to 50\%.
Unfortunately, the rumor ended with CLEO 1999 
direct $a_{\rm CP}$ results in 5 modes,
all consistent with zero with errors $\sim 20\%$.

Along a different line,
we compiled\cite{CHH} measurables which could test
$b_L s_R \gamma$ couplings that 
would definitely indicate New Physics:
mixing dependent and
direct CP violation in $B\to K^*\gamma$,
and $\Lambda$ polarization
in $\Lambda_b \to \Lambda\gamma$ decay.

It turns out that $b\to d$ penguins
may be more promising and accessible in a couple of years.
Let us illustrate the interplay of flavor symmetries and SUSY
in a relatively extreme case.\cite{CH}
For an underlying Abelian horizontal symmetry
for observed quark mass and mixing hierarchy pattern,
1-3 and 2-3 mixings in $d_R$ sector are naturally the largest,
and likewise for $\tilde d_{R}$ with SUSY.
Cabibbo (1-2) mixing must come from up sector
because of $\varepsilon$ and $\varepsilon^\prime$ constraints.
One would have to accept TeV scale squarks and gluinos,
but

\noindent \phantom{x} $\star$
 $B_d$-$\bar B_d$, $B_s$-$\bar B_s$ and $D^0$-$\bar D^0$
	mixing all have common source\\
 \phantom{xxxx} 
  $\Longrightarrow$ strength and CP patterns different from SM!

\noindent \phantom{x} $\star$
 $B\to \rho\gamma$, $\omega\gamma$:
   More detectable ({\it vs.} $K^{*0}\gamma$) mixing-dep. CP asymmetries.\\
Thus, the B system probes New Physics even 
when elusive at colliders.

\section{The Next 3 Years: A Sampler}

B Factories will reach $10^{34}$, 
and Tevatron Run-II will have $20\times$ Run-I data
during 2001-2003.
Clearly the next 3 years would be 
even more exciting than the last 3: 
\fbox{$> 10^8\ B\bar B$ by 2003}, 
and comparable jump at Tevatron!

What could be revealed?
\begin{description}
\item[\underline{\boldmath{$\gamma$ (or $\phi_3$)} Program}]:
  Model-indep. methods ({\it e.g.} $R_*$)
   vs. ``Global Fit". \\
  Question of model indep. {\it vs.} stat. power;
   both can be PQCD-improved.
\item[\underline{\boldmath{$\alpha$ (or $\phi_2$)} Program}]:
  $\alpha \equiv \pi -(\beta+\gamma)$\\
  Since $\bar B^0\to \pi^+\pi^-$ (direct: $\gamma$ in $V_{ub}^*$)
  and $\bar B^0 \to B^0\to \pi^+\pi^-$ (mixing:~$\beta$)
  interfere, mixing-dep. CP probes $\alpha$.
  However, ``P-pollution" severe.\\
  \phantom{x}$\Longrightarrow$ Two Paths:
   \ \ $\pi^+\pi^-$ and $\pi^0\pi^0$ plus isospin analysis\\
   \phantom{AABBCcXXYYZZ}
    $\pi^+\pi^-\pi^0$ ($\rho^\pm\pi^\mp$, $\rho^0\pi^0$)
     Dalitz plot analysis\\
  This is an area where a lot of {\it new} development is expected.
\item[\underline{\boldmath{$\eta^\prime$} Program}]:
 Confirm CLEO, esp. inclusive! (Maybe 
 				 $B\to Glueball + K$?).
\item[\underline{Direct CP Asymmetries}]:
 $a_{\rm CP}$ sensitivity down to few \%.\\
  $\Longrightarrow$ {\it New impact and info on/for theory.}
      (Perhaps just testing FSI...?)
\item[\underline{\boldmath{$\beta$}--crosscheck/NP probe}]:
			 Mixing-dep. CP study in\\
 $\star$ $B\to \phi K_S$:
\footnote{
Belle reported at Osaka a large $\phi K^+$ signal $> 10^{-5}$,
in some conflict with a smaller number reported by CLEO.
Note that the new CLEO number is
above their previous upper limit.
}
     Pure P ($s\bar ss$) w/o SM phase \ $\Rightarrow \,
	\beta_{\phi K_S} \neq \beta_{\psi K_S}$ means NP.\\
  $\star$ $B\to \eta^\prime K_S$: \,
  	Not pure-P (has T) but possibility of NP source.
\item[\underline{EWP \& \boldmath{$b\to d\gamma$}}]:\\
 $\star$ $B\to K^{(*)}\ell^+\ell^-$ {\it will} appear:
   $A_{\rm FB}$ via $\gamma$-$Z$ interference at $m_b$ scale!\\
 $\star$ $B\to \rho\gamma$, $\omega\gamma$
  ($\pi^+\pi^-(\pi^0)$: vertex) vs. $K^{*0}\gamma$
	       ($K^{*0}\to K_S\pi^0$ no vertex)\\
   \phantom{x} $\Longrightarrow$ Good mixing-dep. CP probe: \
 Nonzero $a_{\rm CP}^{\rm mix} \equiv$ New Physics.
\item[\underline{D mixing}]: Confirm CLEO/FOCUS?
 $\Delta m_D \neq 0$ implies New Physics.
\item[\underline{Charmless Rare Baryons}]:\cite{HS} 
 $B\to \eta^\prime\bar\Lambda p;\; \gamma\bar\Lambda p$?\\
   CLEO just reported\cite{Yeldon} $B \to D^{*-} p\bar n$,
   			 $D^{*-} p\bar p\pi \sim 10^{-3}
			  \ltap D^*\pi,\ D^*\rho$!\\
   \phantom{} $\Longrightarrow$ 
   $\eta^\prime\bar\Lambda p;\; \gamma\bar\Lambda p \sim 10^{-5}
    \gg \bar \Lambda p$ plausible,
     {\it could be first charmless baryon!}\\
     \phantom{XXj} $\Lambda \to p\pi$ self-analyze spin:
       {\it probe $B\to \eta^\prime,\; \gamma$ dynamics
            (and CP/T)}.
\end{description}

\section{Conclusion}

We have witnessed the riches of rare B decays 
from the past on weak dynamics, weak phases, 
new physics, and strong interaction.
The highlights have been:
$B\to K^*\gamma$, $K+n\pi +\gamma$;
$\eta^\prime K$, $\eta^\prime +K+n\pi$;
$K\pi/\pi\pi$.

The timeline is illustrated as follows:

\vskip0.2cm
\begin{tabular}{ccll}
Time   & \# $B\bar B$  & Discovery & Significance \\
\hline
1986       &      $10^5$   & $B$-$\bar B$ Mixing & !! \\
1993       & few $10^6$ & $B\to K^*\gamma$, $X_s\gamma$
				& EMP \\
1997--2000 &      $10^7$
	   & $B\to \eta^\prime K$, $\eta^\prime X_s$, $K\pi$,
	     $\phi K$\ldots  & Strong $b\to s$ P \\
     &     & $B\to \rho\pi$, $\omega\pi$, $\pi\pi$
				& $b\to u$ T \& P \\
2000--2003 &      $10^8$   & $\phi_1/\beta$, $\phi_3/\gamma$,
			     $\phi_2/\alpha$ (?)
				& Unitarity $\triangle$ \\
     &     & Direct $a_{\rm CP}$ & FSI, or \ldots? \\
     &     & $B\to K^{(*)}\ell^+\ell^-$  & EWP \\
     &     & $B\to \rho\gamma,\ \omega\gamma$   & $b\to d$ EMP \\
     &     & \ldots\ldots	& New Physics!?
\end{tabular}

With advent of B Factories,
we expect an order or more jump in number of $B\bar B$'s, boosted.
Detectors now have good PID plus vertexing.
Boom time lies ahead --- Era of BaBar/Belle/CLEO competition,
and with Tevatron.

\end{document}